\documentclass[twoside]{mhd}
\usepackage{graphicx}
\usepackage{bm}
\mhdhead{40}{1}{1}	

\title{TOWARDS A PRECESSION DRIVEN DYNAMO EXPERIMENT}

\author{F. Stefani\inst{1}, T. Albrecht\inst{2}, G. Gerbeth\inst{1}, 
A. Giesecke\inst{1}, T. Gundrum\inst{1},\\
J. Herault\inst{1}, C. Nore\inst{3}, C. Steglich\inst{1}}

\institute{Helmholtz-Zentrum Dresden-Rossendorf, 
P.O. Box 510119, D-01314 Dresden, Germany\\ 
\and
Dep.~of Mech.~and Aerospace Engineering, Monash University, 
VIC 3800, Australia\\
\and
LIMSI-CNRS/B\^{a}timent 508, BP 133, 
91403 Orsay cedex, France et Universit\'{e} 
Paris-Sud, 91405 Orsay cedex, France}

\begin{document}
\maketitle


\begin{abstract}\noindent
The most ambitious project within the DREsden 
Sodium facility for DYNamo and thermohydraulic studies 
(DRESDYN) at Helmholtz-Zentrum Dresden-Rossendorf (HZDR)
is the set-up of a precession-driven 
dynamo experiment. After discussing the scientific 
background and some results of water pre-experiments 
and numerical predictions, we focus on the numerous 
structural and design problems of the machine. We also 
outline the progress of the building's construction, and 
the status of some other experiments that are planned in 
the framework of DRESDYN. 
\end{abstract}


\section*{Introduction}
Pioneered by the Riga \cite{RIGA} and Karlsruhe \cite{KARLSRUHE} 
liquid sodium experiments, the last fifteen years have seen 
significant progress in the experimental study of the dynamo 
effect and of related magnetic instabilities, such as the 
magnetorotational instability (MRI) \cite{SISAN,MRI_PRE,SEILMAYER2} 
and the kink-type 
Tayler instability (TI) \cite{SEILMAYER1}. A milestone 
was the observation of magnetic field reversals in the VKS experiment 
\cite{BERHANU} which has spurred renewed interest in simple models to 
explain the corresponding geomagnetic phenomenon \cite{PETRELIS}. This is 
but one example for the fact that liquid metal experiments, though never 
representing perfect models of specific cosmic bodies, can indeed 
stimulate geophysical research.

One of the pressing questions of geo- and astrophysical 
magnetohydrodynamics concerns the energy source of different cosmic 
dynamos. While thermal and/or compositional buoyancy is considered 
the favourite candidate, 
precession has long been discussed as a complementary energy source of 
the geodynamo \cite{MALKUS,GANS,VANYO,TILGNER,SHALIMOV}, 
in particular 
at an early stage of Earth's evolution, prior to the formation 
of the solid core. Some influence of 
orbital parameter variations can also be guessed from paleomagnetic 
measurements 
that show an impact of the 100 kyr Milankovic cycle of the Earth's 
orbit eccentricity on the reversal statistics of the geomagnetic field 
\cite{CONSOLINI}. 
Recently, precessional driving has also been 
discussed in connection with the 
generation of the lunar magnetic field \cite{DWYER}, and with dynamos in 
asteroids \cite{FU}.

Whilst, therefore, an experimental validation of precession driven dynamo 
action appears very attractive, the constructional effort and safety 
requirements for its realization are tremendous. In this paper, we 
outline the present state of the preparations of such an experiment, 
along with giving an overview of further liquid sodium experiments 
that are planned within 
DREsden Sodium facility for DYNamo and thermohydraulic studies 
(DRESDYN) at Helmholtz-Zentrum Dresden-Rossendorf (HZDR).

\begin{figure}[t]
\begin{center}
\unitlength=\textwidth
\includegraphics[width=0.99\textwidth]{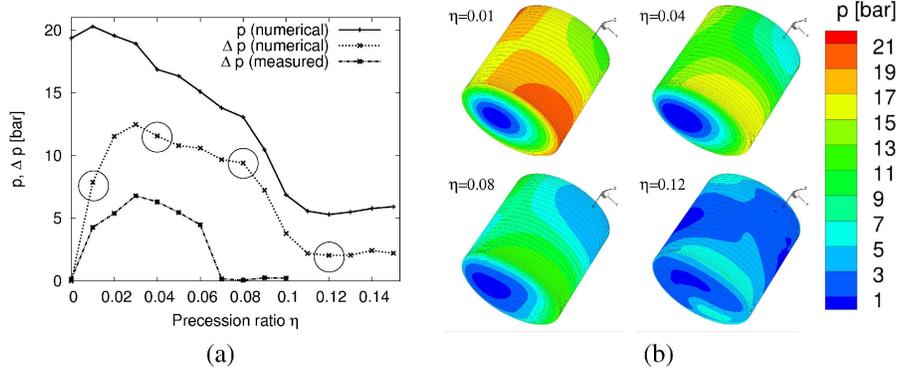}
\caption{Pressure values for a precessing cylinder with $90^{\circ}$ angle 
between rotation and precession axes, in dependence on the precession ratio $\eta$, 
when scaled to the dimensions of the 
large sodium device. (a) Maximum pressure $p$ 
(numerically determined at $Re=6680$) 
and maximum 
pressure difference $\Delta p$ (numerical and experimental). 
(b) Pressure distribution at the rim of the cylinder 
(numerical) for 4 specific precession ratios.}
\end{center}
\end{figure}

\section{To B or not to B}

Compared to the flow structures underlying the Riga, 
Karlsruhe and the Cadarache VKS experiment, the dynamo action of 
precession driven flows is not well understood. Recent dynamo 
simulations in spheres \cite{TILGNER}, cubes \cite{KRAUZE}, and 
cylinders \cite{NORE} were typically carried 
out at Reynolds numbers $\rm{Re}$ of a few thousand, and with magnetic 
Prandtl numbers $Pm$ not far from 1. Under these conditions, dynamo 
action in cubes and cylinders was obtained at magnetic Reynolds 
numbers $\rm{Rm}:=\mu_0 \sigma R^2 \Omega_{rot}$ of around 700 
($\mu_0$ is the magnetic permeability constant, $\sigma$ the conductivity, 
$R$ the radius or half sidelength in case of a cube, 
$\Omega_{\rm{rot}}$ is the angular velocity of rotation), which
is indeed the value our experiment is aiming at. Yet, there are 
uncertainties about this value, mainly because 
numerical simulations fail at 
realistic Reynolds numbers.
For this purpose, a 1:6 scaled water experiment 
(described in \cite{MAHYD}) has been set-up and  
used for various flow measurements, complementary 
to those done at the ATER experiment 
in Paris-Meudon \cite{JACQUES,MOUHALI}. 

So far, we have achieved some qualitative, 
though not quantitative, agreement of the dominant flow structures 
between experiment and numerics for precessing cylindrical flows,
at least for angles between precession and rotation axis not very
far from 90$^{\circ}$. 
Basically, at low precession 
ratios $\eta:=\Omega_{\rm{prec}}/\Omega_{\rm{rot}}$ 
($\Omega_{\rm{prec}}$ is the angular velocity of precession) the  
flow is dominated by the first ($m=1$) 
Kelvin mode. Approximately at $\eta=0.03$, higher azimuthal 
modes appear which start to draw more and more energy from the 
forced $m=1$ Kelvin mode.
Still laminar, this regime breaks down 
suddenly at $\eta \sim 0.07$ (details depend 
on the aspect ratio of the cylinder, the angle between rotation and 
precession axis, and the Reynolds number). 
We identified two global features by which this 
laminar-turbulent transition can be easily characterized. The first 
one is the energy dissipation, measurable by the motor power of the 
rotating cylinder (see \cite{MAHYD}). The second one is the maximum 
pressure difference between opposite points on the side wall of the 
cylinder. Figure~1a 
shows the maximum pressure $p$ (numerically determined at $Re=6680$) and 
the maximum pressure difference $\Delta p$ (numerically determined at 
$\rm{Re}=6680$ and experimentally at $\rm{Re}=1.6 \times 10^6$), with all 
values up-scaled 
to the dimensions of the large machine. The right end-point, at 
$\eta \approx ~0.07$, of the parabola-like part of the 
experimental $\Delta p$ 
curve, marks 
the sudden transition from  the laminar to the turbulent regime. The 
corresponding numerical $\Delta p$ curve is qualitatively similar, but 
shows significant quantitative deviations, in particular a shift of the
transition point towards higher values of $\eta$. Note also that
the maximum value of $\Delta p$ appears approximately at $\eta=0.03$
from where on the higher $m$ modes are increasingly 
fed by the forced $m=1$ mode.

The effect of the Reynolds number on the various 
transitions found in our experiment is shown in Fig. 2. It shows also 
a first transition (diamonds) from a stable, 
$m=1$-Kelvin-mode dominated
flow to a more unstable flow comprising also higher $m$-modes.
The two upper lines indicate the laminar-turbulent transition, 
either coming from the laminar side (circles), or from the 
turbulent side
(squares). The difference between the two lines indicates a 
hysteresis. 
The Reynolds number only weakly affects the transitions (an empirical 
curve fit for laminar-turbulent transition corresponds to 
$\sim \rm{Re}^{-0.067}$), raising hopes that the large machine 
might not behave 
dramatically different.

\begin{figure}[t]
\begin{center}
\unitlength=\textwidth
\includegraphics[width=0.8\textwidth]{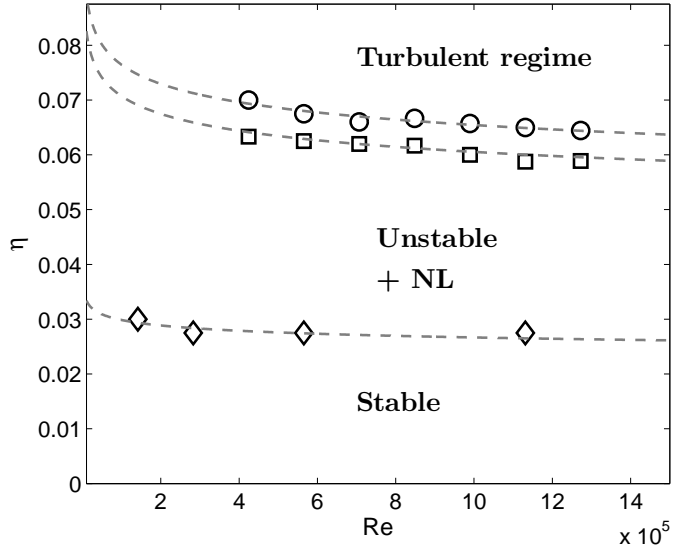}
\caption{Effect 
of Reynolds number on the transition points between different flow
regimes
for an angle between rotation and
precession axes of 90$^{\circ}$. 
Diamonds mark the 
first bifurcation from the pure forced $m=1$ 
Kelvin mode. Circles mark the boundary between the unstable, 
non-linear (NL) regime and the turbulent regime. Squares 
mark the same boundary but when the system 
returns from the turbulent regime (hysteresis). The 
curve fits of the upper two 
data sets correspond to $\sim \rm{Re}^{-0.067}$ in either case,
the curve fit of the lower data set corresponds to 
$\sim \rm{Re}^{-0.049}$.}
\end{center}
\end{figure}

Interestingly, an intermediate regime characterized by the 
occurrence of a few medium-sized cyclones has been observed 
at the ATER experiment in Paris-Meudon \cite{MOUHALI}. So 
far, these vortex-like 
structures could not be identified at our water experiment. 
A 3-D, volumetric particle image velocimetry system currently 
being commissioned could ultimately provide a helicity distribution, 
which in turn could be fed to dynamo simulations.
In general, we 
expect more conclusive dynamo predictions, in particular for 
the cyclonic and the turbulent regime, from a close interplay 
of water test measurements and advanced numerical simulations.

\begin{figure}[t]
\begin{center}
\unitlength=\textwidth
\includegraphics[width=0.8\textwidth]{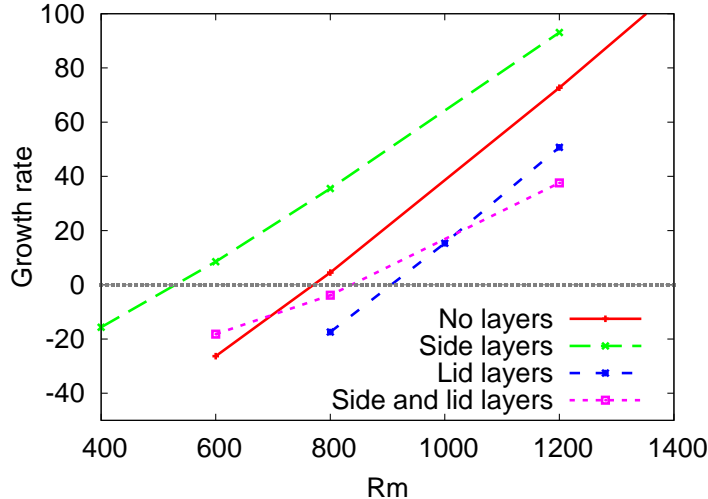}
\caption{Growth rates of the magnetic field energy 
(measured in units of the diffusion time scale 
$\mu \sigma R^2$) in dependence on 
specific  electrical boundary 
conditions. The values are for $\rm{Re}=1200$ and precession ratio 
$\eta=0.15$. The thickness of either layer type 
is taken as $0.1 R$, with the same conductivity as the fluid.}
\end{center}
\end{figure}

Up to present, dynamo action for precessing cylindrical 
flows has been confirmed in nonlinear simulations of the 
MHD equations for the 
case $\rm{Pm}:=\mu_0 \sigma \nu \sim 1$ ($\nu$ 
is the kinematic viscosity) and 
$\eta=0.15$ \cite{NORE}. Yet, the 
critical $\rm{Rm}$ depends on the 
specific electrical boundary conditions (see Fig. 3), 
with a surprisingly low 
optimum value of 550 for the case of electrically conducting 
side layers and insulating lid layers (actually, this finding 
has led us to consider an inner copper layer attached to the outer 
stainless steel shell of the dynamo vessel).
Albeit this low critical $\rm{Rm}$ is encouraging, the question 
of self-excitation in a real precession experiment is far 
from being settled. Further simulations at lower $\rm{Pm}$ have 
led to an increase of the critical $\rm{Rm}$, and for lower values 
of $\eta$ dynamo action has not been confirmed yet 
(see also \cite{GIESECKE}).

\section{Status of preparations}

In contrast to previous dynamo experiments, the precession 
experiment has a higher degree of homogeneity since it 
lacks impellers or guiding blades and any magnetic material
(the latter seems to play a key role for the 
functioning of the VKS dynamo \cite{GIESECKE_PRL}). 
The central precession module encases 
a sodium filled cylindrical volume of 2 m diameter and the same 
height (Figure 4). For this volume, we aim at reaching a rotation 
rate of 10 Hz (to obtain $\rm{Rm}=700$), and a precession rate of 1 Hz 
(to cover both the laminar and the turbulent flow regimes). 
With total gyroscopic torques of up 
to $8 \times 10^6$ Nm, we are in many respects  
at the edge of technical feasibility, so that much optimization work 
is needed to enable safe operation of the machine.

The complicated simultaneous rotation around two axes poses 
several challenges: filling and emptying procedures, heating 
and cooling methods, and handling of thermal expansion. A 
decision was made in favor of a slightly enlarged vessel, 
comprising two conical end-pieces that serve, first, for a 
well-defined filling and emptying procedure at 43$^\circ$ vessel 
tilting, and, second, for hosting two bellows which will
compensate the thermal expansion of the liquid sodium.

\begin{figure}[t]
\begin{center}
\unitlength=\textwidth
\includegraphics[width=0.99\textwidth]{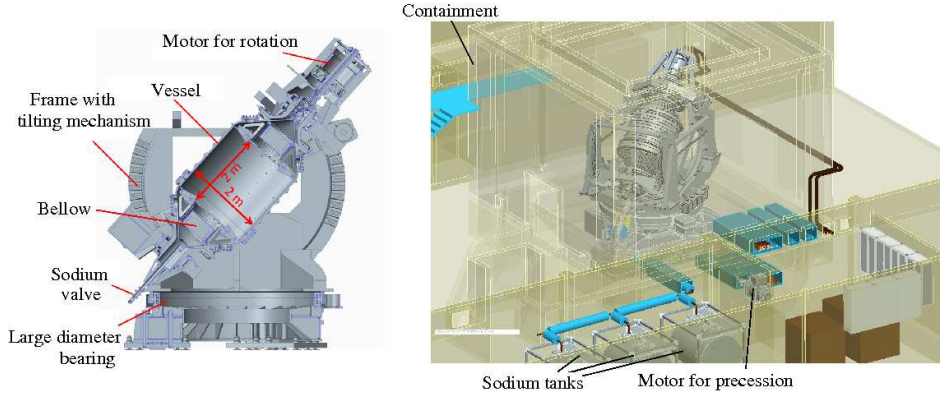}
\caption{Present status of the design of the precession experiment.}
\end{center}
\end{figure}

Having defined this basic structure of the central vessel, 
much effort was, and still is, devoted to the optimization 
of the shell. A shell thickness of around 3 cm is needed 
anyway to withstand the centrifugal pressure of 20 bar in 
case of pure rotation (see Fig. 1). For increasing precession ratio,
this total pressure decreases, but is complemented 
by a pressure vacillation $\Delta p$ due to the gyroscopic forces. 
In addition to those mechanical stresses, we also have to 
consider thermal stresses caused by the temperature 
difference over the shell when the dynamo is cooled by a strong 
flow of air. A particular problem is the high mechanical stress
in the holes for the measurement flanges.

The next step is the design of the bearings and of a frame 
that allows to choose different angles between rotation and 
precession axis. Finding appropriate roller bearings for the 
vessel turned out to be extremely challenging, mainly because 
of the huge gyroscopic torque. It is the same gyroscopic torque 
that also requires a very stable basement (Fig. 5a,b), supported 
by seven pillars, each extending 22 m into the bedrock. The 
precession experiment itself is embedded in a containment 
(Fig. 4, Fig. 5b,c), 
preventing the rest of the building from the consequences of 
possible sodium leaks. Since the double rotation cannot be 
stopped quickly 
in case of an accident, this containment is the only chance of 
preventing jets that would spill out of a potential leak from 
perfectly covering all surrounding areas with burning sodium. 
For such accidents, the containment can be flooded with 
argon, which is stored in liquid form.

\begin{figure}[t]
\begin{center}
\unitlength=\textwidth
\includegraphics[width=0.99\textwidth]{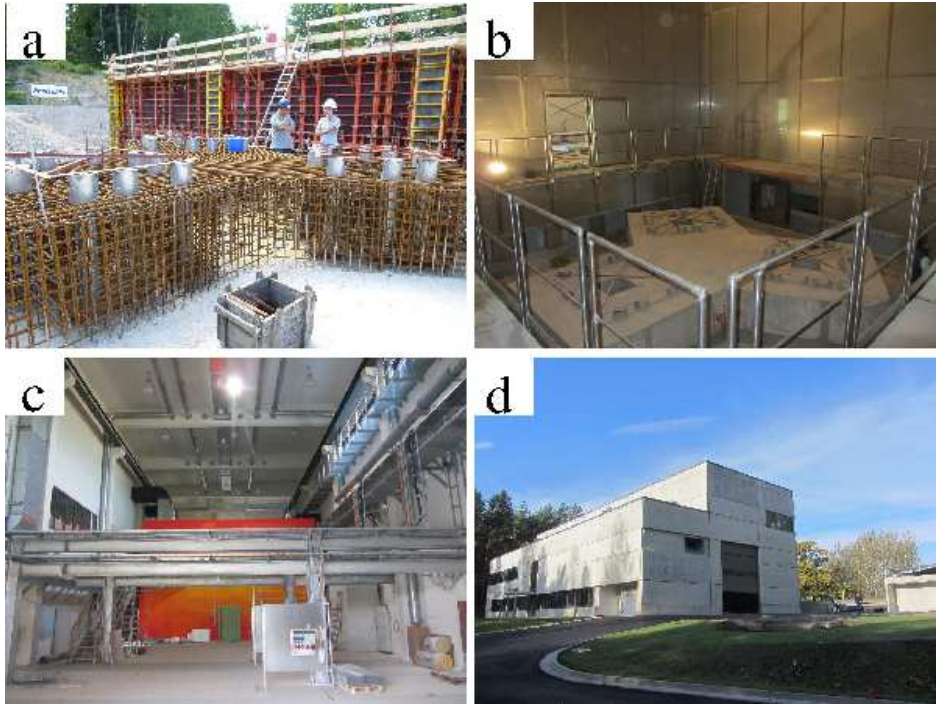}
\caption{The DRESDYN building. (a) Construction of the ferro-concrete
tripod for the precession experiment (August 2013). This separate basement is 
supported by 7 pillars with a depth of 22 m. (b) The tripod 
situated in the containment.
(c) View into the experimental hall, with the (coloured) 
containment in the background. (d) The DRESDYN building as of October 2014.
}
\end{center}
\end{figure}

\section{DRESDYN---What else is it good for?}

Given the significant investment needed for the very 
precession experiment and the infrastructure to support it, we 
have combined this specific installation with creating a general 
platform for further liquid metal experiments. 

A second  experiment 
relevant to geo- and astrophysics will investigate 
different combinations of the 
MRI and the current-driven TI (see Fig. 6). Basically, the 
set-up is a 
Taylor-Couette experiment with 2 m height, an inner 
radius $R_{\rm{in}}=20$ cm and an outer radius 
$R_{\rm{out}}=40$ cm. 
Rotating the inner cylinder at up to 20 Hz we plan 
to reach an $\rm{Rm}$ of around 40, while the axial magnetic field $B_z$ 
will correspond to a Lundquist number 
$S:=\mu \sigma R_{in} B_{z}/\sqrt{\mu_o \rho}$ 
of 8. Both values are about twice their 
respective critical values \cite{SHALYBKOV} for the standard 
version of MRI 
(with only an axial magnetic field applied). 

Below those critical values, 
we plan to investigate how the helical version 
of MRI approaches the limit of standard MRI \cite{HOLLERBACH,KIRILLOV_APJ}. 
To this end, 
we will use a strong central current, as it was already done in 
the PROMISE experiment \cite{MRI_PRE,SEILMAYER2}. This insulated 
central current can 
be supplemented by another axial current, guided through the 
(rotating) liquid sodium, which will then allow to investigate 
arbitrary combinations of MRI and TI. Recent theoretical 
studies \cite{KIRILLOV_PRL,KIRILLOV_JFM} 
have shown that even a slight addition of current 
through the liquid would extend the range of application of the 
helical and azimuthal MRI to Keplerian flow profiles.

\begin{figure}[t]
\begin{center}
\unitlength=\textwidth
\includegraphics[width=0.8\textwidth]{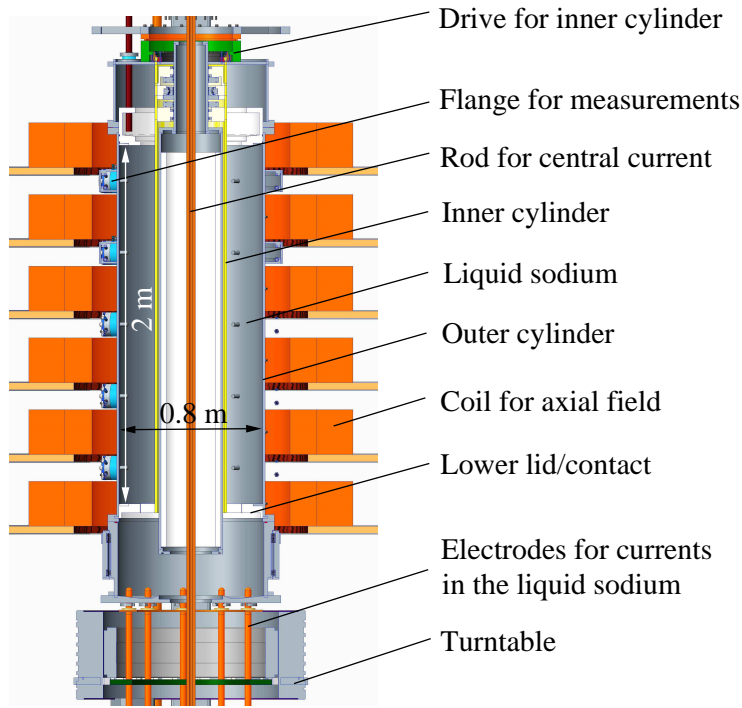}
\caption{Present status of the design of the combined 
MRI/TI experiment.}
\end{center}
\end{figure}

The TI will also play a central role in a third experiment in 
which different flow instabilities in liquid metal batteries 
(LMB) will be studied. LMB's consist of three self-assembling 
liquid
layers \cite{SADOWAY}, an alkali or earth-alkali metal (Na, Mg), 
an electrolyte, 
and a metal or half-metal (Bi, Sb). In order to be competitive, 
LMB's have to be quite large, so that charging 
and discharging currents in the order of some kA are to be expected. 
Under those conditions, TI and interface 
instabilities must be carefully avoided \cite{NORBERT1,NORBERT2}.

Another installation is an In-Service-Inspection (ISI) experiment 
for various studies related to safety aspects of sodium fast 
reactors (SFR). In this context we also intend to investigate 
experimentally the 
impact of magnetic materials  (e.g. ODS steels) 
on the mean-field coefficients in 
spiral flow configurations (see \cite{GIESECKE2}), and its 
consequences for the possibility of magnetic-field 
self-excitation in the cores of SFR's.

The construction of the DRESDYN building is well advanced. 
Figure 5c,d illustrate the status of the construction as of 
October 2014. The interior construction is expected to be finalized 
in 2015. Thereafter, installation of the various experiments 
can start. It goes without saying that both the precession and 
the MRI/TI experiment will be tested with water first, before we 
can dare to run them with liquid sodium.

\section{Conclusions}

We have discussed the motivation behind, and the
concrete plans for a number of experiments to be set-up in the 
framework of DRESDYN. The new building and the essential parts of the
experiments are expected to be ready in 2015. 
Apart from hosting the discussed experiments, 
DRESDYN is also meant as a general platform for further large-scale 
experiments, basically but not exclusively with liquid sodium. 
Proposals for  such experiments are, therefore, highly welcome.

\section*{Acknowledgments}
This work was supported 
by Helmholtz-Gemeinschaft
Deutscher Forschungszentren (HGF) in frame  of 
the Helmholtz Alliance ''Liquid metal technologies'' (LIMTECH), and
by Deutsche Forschungsgemeinschaft 
(DFG) under grant STE 991/1-2.
We thank Jacques L\'eorat for his proposal 
and encouragement to 
set-up a precession driven dynamo.
Fruitful discussions with Andreas Tilgner are gratefully 
acknowledged.



\begin{thebibliography}{50}

\bibitem{RIGA}
{\sc A. Gailitis et al}.
\newblock {Detection of a flow induced magnetic field eigenmode in the Riga dynamo facility}.
\newblock {\it {Phys. Rev. Lett.}\/}, vol.~84 (2000), pp.~4365-4368.

\bibitem{KARLSRUHE}
{\sc R. Stieglitz and U M\"uller}.
\newblock {Experimental demonstration of an homogeneous two-scale dynamo}.
\newblock {\it {Phys. Fluids}\/}, vol.~13 (2001), pp. 561--564.

\bibitem{SISAN}
{\sc D.R. Sisan et al}.
\newblock {Experimental observation and characterization of the magnetorotational instability}.
\newblock {\it {Phys. Rev. Lett.}\/}, vol.~93 (2004), Art. No. 114502.


\bibitem{MRI_PRE}
{\sc  F. Stefani et al}.
\newblock {Helical magnetorotational instability in 
a Taylor-Couette flow with strongly reduced Ekman pumping}.
\newblock {\it {Phys. Rev. E}\/} vol.~97 (2006), Art. No. 184502.

\bibitem{SEILMAYER2}
{\sc  M. Seilmayer et al}.
\newblock {Experimental evidence for non-axisymmetric magnetorotational instability in an azimuthal magnetic field }.
\newblock {\it {Phys. Rev. Lett.}\/} vol.~113 (2014), Art. No. 024505.


\bibitem{SEILMAYER1}
{\sc  M. Seilmayer et al}.
\newblock {Experimental evidence for a transient Tayler instability in a cylindrical liquid-metal column}.
\newblock {\it {Phys. Rev. Lett.}\/} vol.~108 (2012), Art. No. 024501.


\bibitem{BERHANU}
{\sc  M. Berhanu et a}.
\newblock {Magnetic field reversals in an experimental turbulent dynamo}.
\newblock {\it {Europhys. Lett.}\/} vol.~77  (2007), Art. No. 59001.



\bibitem{PETRELIS}
{\sc  F. Petrelis, S. Fauve, E. Dormy, and J.P. Valet}.
\newblock {Simple mechanism for reversals of Earth's magnetic field}.
\newblock {\it {Phys. Rev. Lett.}\/} vol.102  (2009), Art. No. 144503.


\bibitem{MALKUS}
{\sc W.V.R. Malkus}.
\newblock {Precession of Earth as cause of geomagnetism}.
\newblock {\it {Science}\/}, vol.~160 (1968), pp.~259--264.

\bibitem{GANS}
{\sc R.F. Gans}.
\newblock {On hydromagnetic precession in a cylinder}.
\newblock {\it {J. Fluid Mech.}\/}, vol.~45 (1970), pp.~111--130.

\bibitem{VANYO}
{\sc J.P. Vanyo}.
\newblock {Core-mantle relative motion and coupling}.
\newblock {\it {Geophys. J. Int.}\/}, vol.~158 (2004), pp.~470--478.

\bibitem{TILGNER}
{\sc A. Tilgner}.
\newblock {Precession driven dynamo}.
\newblock {\it {Phys. Fluids}\/}, vol.~17 (2005), Art. No.~034104.

\bibitem{SHALIMOV}
{\sc S.L. Shalimov}.
\newblock {On the precession driven geodynamo}.
\newblock {\it {Izvestiya, Phys. Sol. Earth}\/}, vol.~42 (2005), pp. 460--466.

\bibitem{CONSOLINI}
{\sc G. Consolini and P. De Michelis}.
\newblock {Stochastic resonance in geomagnetic polarity reversals}.
\newblock {\it {Phys. Rev. Lett.}\/}, vol.~90 (2003), Art. No.~058501.



\bibitem{DWYER} 
{\sc C.A. Dwyer, D.J. Stevenseon, and F. Nimmo}.
\newblock {A long-lived lunar dynamo driven by continuous mechanical stirring}.
\newblock {\it {Nature}\/}, vol.~220 (2012) pp.~47-61.


\bibitem{FU} 
{\sc R.R. Fu et al}.
\newblock {An ancient core dynamo in asteroid Vesta}.
\newblock {\it {Science}\/}, vol.~338 (2013) pp.~238-241.


\bibitem{KRAUZE}
{\sc A. Krauze}.
\newblock {Numerical modeling of the magnetic field generation in a precessing cube
with a conducting melt}.
\newblock {\it {Magnetohydrodynamics}\/}, vol.~46 (2010), pp.~271-280.


\bibitem{NORE}
{\sc C. Nore, J. L\'eorat, J.L. Guermond, and F. Luddens}.
\newblock {Nonlinear dynamo action in a precessing cylindrical container}.
\newblock {\it {Phys. Rev. E}\/}, vol.~84 (2011), Art. No.~016317.


\bibitem{MAHYD}
{\sc F. Stefani et al.}.
\newblock {DRESDYN - A new facility for MHD experiments with liquid sodium}.
\newblock {\it {Magnetohydrodynamics}\/}, vol.~46 (2010), pp.~271-280.

\bibitem{JACQUES}
{\sc J. L\'eorat}.
\newblock {Large scales features of a flow driven by precession}.
\newblock {\it {Magnetohydrodynamics}\/}, vol.~42 (2006), pp.~143-151.



\bibitem{MOUHALI}
{\sc W. Mouhali, T. Lehner, J. L\'eorat, and R. Vitry}.
\newblock {Evidence for a cyclonic regime in a precessing cylindrical container}.
\newblock {\it {Exp. Fluids}\/}, vol.~53 (2012), 1693-1700.

\bibitem{GIESECKE}
{\sc A. Giesecke, T. Albrecht, G. Gerbeth, T. Gundrum, and F. Stefani}.
\newblock {Numerical simulation for the DRESDYN precession dynamo}.
\newblock {\it {Magnetohydrodynamics}\/}, this issue (2015).



\bibitem{GIESECKE_PRL}
{\sc A. Giesecke, F. Stefani, and G. Gerbeth}.
\newblock {Role of soft-iron impellers on the mode selection 
in the von K\'{a}rm\'{a}n-sodium dynamo experiment},
\newblock {\it {Phys. Rev. Lett.}\/}, vol.~104 (2010), Art. No. 044503.


\bibitem{SHALYBKOV}
{\sc G. R\"udiger, M. Schultz, and D. Shalybkov}.
\newblock {Linear magnetohydrodynamic Taylor-Couette instability for liquid sodium}.
\newblock {\it {Phys. Rev. E}\/}, vol.~67 (2003), Art. No.~046312.

\bibitem{HOLLERBACH}
{\sc R. Hollerbach and G. R\"udiger}.
\newblock {New type of magnetorotational instability in cylindrical Taylor-Couette flow
}.
\newblock {\it {Phys. Rev. Lett.}\/}, vol.~95 (2005), Art. No.~124501.

\bibitem{KIRILLOV_APJ}
{\sc O.N. Kirillov and F. Stefani}.
\newblock {On the relation of standard and helical magnetorotational instability}.
\newblock {\it {Astrophys. J.}\/}, vol.~712 (2010), pp. 52-68.


\bibitem{KIRILLOV_PRL}
{\sc O.N. Kirillov and F. Stefani}.
\newblock {Extending the range of the inductionless magnetorotational instability}.
\newblock {\it {Phys. Rev. Lett.}\/}, vol.~111 (2013), Art. No.~061103.


\bibitem{KIRILLOV_JFM}
{\sc O.N. Kirillov, F. Stefani, and Y. Fukumoto}.
\newblock {Local instabilities in magnetized rotational flows: a short-wavelength approach}.
\newblock {\it {J. Fluid Mech.}\/}, in press (2014); arXiv:1401.8276.


\bibitem{SADOWAY}
{\sc H. Kim et al}.
\newblock {Liquid metal batteries: Past, present, and future}.
\newblock {\it {Chem. Rev.}\/}, vol.~113 (2013), 2075-2099.

\bibitem{NORBERT1}
{\sc N. Weber, V. Galindo, F. Stefani, T. Weier, and T. Wondrak}.
\newblock {Numerical simulation of the Tayler instability in liquid metals}.
\newblock {\it {New J. Phys.}\/}, vol.~15 (2013), Art. No.~043034.

\bibitem{NORBERT2}
{\sc N. Weber, V. Galindo, F. Stefani, and T. Weier}.
\newblock {Current-driven flow instabilities in large-scale liquid metal batteries, and how to tame them}.
\newblock {\it {J. Power Sources}\/}, vol.~265 (2014), 166-173.



\bibitem{GIESECKE2}
{\sc A. Giesecke, F. Stefani, and G.Gerbeth}.
\newblock {Magnetic material in mean-field dynamos driven by small scale helical flows}.
\newblock {\it {New J. Phys.}\/}, vol.~16 (2014), Art. No.~073034.








\end{thebibliography}

\newcommand{\noopsort}[1]{}

\lastpageno

\end{document}